\documentclass[]{aastex63}

\shorttitle{The Molecular Exoskeleton of NGC 3132}
\shortauthors{Kastner et al.}

\begin{document}

\title{The Molecular Exoskeleton of the Ring-like Planetary Nebula NGC 3132}

\correspondingauthor{Joel Kastner}
\email{jhk@cis.rit.edu}

\author[0000-0002-3138-8250]{Joel H. Kastner}
\affiliation{Center for Imaging Science, 
  Rochester Institute of Technology, Rochester NY 14623, USA; jhk@cis.rit.edu}
\affiliation{School of Physics and Astronomy, 
  Rochester Institute of Technology} 
\affiliation{Laboratory for Multiwavelength Astrophysics, 
  Rochester Institute of Technology}

\author[0000-0003-1526-7587]{David J. Wilner}
\affiliation{Center for Astrophysics, Harvard \& Smithsonian, 60 Garden Street, Cambridge, MA 02138-1516, USA}

\author{Paula Moraga Baez}
\affiliation{School of Physics and Astronomy, 
  Rochester Institute of Technology} 
\affiliation{Laboratory for Multiwavelength Astrophysics, 
  Rochester Institute of Technology}

\author{Jesse Bublitz}
\affiliation{Green Bank Observatory, 155 Observatory Road, Green Bank, WV 24944, USA; jbublitz@nrao.edu}

\author{Orsola De Marco}
\affiliation{School of Mathematical and Physical Sciences, Macquarie University, Sydney, New South Wales, Australia}
\affiliation{Astronomy, Astrophysics and Astrophotonics Research Centre, Macquarie University, Sydney, New South Wales, Australia}

\author{Raghvendra Sahai}
\affiliation{Jet Propulsion Laboratory, California Institute of Technology, 4800 Oak Grove Drive, Pasadena, CA 91109, USA}

\author{Al Wootten}
\affiliation{National Radio Astronomy Observatory, Charlottesville, VA 22903, USA}

\begin{abstract}
We present Submillimeter Array (SMA) mapping of $^{12}$CO $J=2\rightarrow 1$, $^{13}$CO $J=2\rightarrow 1$, and CN $N=2\rightarrow 1$ emission from the Ring-like planetary nebula (PN) NGC 3132, one of the subjects of JWST Early Release Observation (ERO) near-infrared imaging. The $\sim$5$''$ resolution SMA data demonstrate that the Southern Ring's main, bright, molecule-rich ring is indeed an expanding ring, as opposed to a limb-brightened shell, in terms of its intrinsic (physical) structure. This suggests that NGC 3132 is a bipolar nebula viewed more or less pole-on (inclination $\sim$15--30$^\circ$).
The SMA data furthermore reveal that the nebula harbors a second expanding molecular ring that is aligned almost orthogonally to the main, bright molecular ring. We propose that this two-ring structure is the remnant of an ellipsoidal molecular envelope of ejecta that terminated the progenitor star's asymptotic giant branch evolution and was subsequently disrupted by a series of misaligned fast, collimated outflows or jets resulting from interactions between the progenitor and one or more companions. 
\end{abstract}
\section{Introduction}

Planetary nebulae  (PNe) are the near-endpoints of stellar evolution for intermediate-mass ($\sim$1--8 $M_\odot$) stars. 
Each PN provides a snapshot of the brief ($\sim$$10^4$ yr) stage in which the outflowing, dusty circumstellar envelope of an asymptotic giant branch (AGB) star is ionized by its newly unveiled core, itself a future white dwarf. 
The resulting $\sim$$10^4$ K circumstellar plasma is a rich source of optical emission lines, forming a classical PN. However, certain PNe retain cold ($<100$ K), dense ($\sim$$10^4$--$10^6$ cm$^{-3}$), massive envelopes of molecular
gas and dust. These PN molecular envelopes are shaped and displaced by fast winds from their exceedingly hot ($\sim$100--200 kK), rapidly evolving central stars, which are also sources of intense UV irradiation of the molecular gas.

The molecule-rich zones of PNe have been detected via IR imaging of H$_2$ rovibrational emission, which reveals shock-heated and/or UV-irradiated molecular gas \cite[e.g.,][]{Webster1988,Zuckerman:1988hb,Kastner1994,Kastner:1996ab}, and by mm-wave spectroscopy of CO rotational emission from far colder and more massive molecular reservoirs within PNe \cite[e.g.,][]{Huggins1996,Huggins2005}. The vast majority of such molecule-rich PNe, most of which are detected in both near-IR H$_2$ and mm-wave CO, are Ring-like or bipolar in structure;  these objects likely constitute a PN class descended from relatively massive progenitor stars \cite[][and references therein]{Kastner:1996ab}. Interferometric observations of such molecule-rich planetary nebulae in the mm-wave regime afford unparalleled opportunities to study their density structures, kinematics, and compositions. The resulting high-resolution molecular line maps of PNe can provide --- among other things --- stringent tests of models of the shaping of such nebulae by collimated outflows from central binary systems as well as insight into the enrichment of the ISM in the products of intermediate-mass stellar nucleosynthesis \cite[e.g.,][]{Kastner2018}. 

Here, we present Submillimeter Array (SMA) mapping of molecular emission from the PN NGC~3132. NGC~3132 is a nearby ($D = 754$ pc), ``Ring-like'' PN that harbors a wide visual binary comprising the central (progenitor) star and an A star companion \citep{Ciardullo1999}. The inner, ionized cavity of NGC~3132 is elliptical in shape, with a major axis of $\sim$40 arcsec (0.15 pc) and an electron density of $n \sim 10^3$\,cm$^{-3}$. The PN's ionization structure and abundances were the subject of a recent optical (VLT/MUSE) spectroscopic mapping study \citep{MonrealIbero2020}.

Like other PNe in its (Ring-like) class \citep{Kastner1994}, NGC~3132 has long been known to harbor a significant mass of molecular gas, as revealed by H$_2$ and CO emission \citep{Storey1984,Sahai:1990fu,Zuckerman1990,Kastner:1996ab}. JWST Early Release Observation (ERO) imaging of NGC 3132 has now revealed the structure of its H$_2$ emission region in unprecedented detail \citep{DeMarco2022}. The JWST H$_2$ images reveal a complex ring system surrounding the central ionized region, as well as a system of arcs within the nebula's extended halo. \cite{DeMarco2022} assert that these structures were most likely sculpted by an unseen companion or companions orbiting within $\sim$60 au of the PN progenitor. Furthermore, the mid-IR JWST (MIRI) images demonstrate the ultra-hot central star has a significant IR excess that most likely emanates from a dusty disk that formed as the result of a close binary interaction, albeit not necessarily with the same companion that generated the H$_2$ ring and arc systems \citep{DeMarco2022,Sahai2023}. 

The H$_2$ emission imaged by JWST only traces the hot ($\sim$1000 K), UV-illuminated and/or shock-excited molecular gas in the nebula, and such hot H$_2$ likely constitutes a small fraction of the total reservoir of molecular gas in NGC 3132. Furthermore, JWST imaging does not provide any information concerning the molecular gas kinematics, such as can be obtained via mm-wave molecular line mapping. However, the only previous such molecular line mapping of NGC 3132 consists of a single-dish $^{12}$CO map obtained with the late SEST facility (beamwidth $\sim$20$''$) well over 30 yr ago \citep{Sahai:1990fu}. These SEST observations revealed strong CO emission from the PN's central ring system that is characterized by expansion at $\sim$15 km s$^{-1}$, with hints of faster ($>$20 km s$^{-1}$) outflows. The only other molecule that has been detected in NGC 3132 thus far (apart from CO and H$_2$) is HCO$^+$ \citep{Sahai1993}.

To establish the distribution, mass, and velocity structure of the molecular gas in NGC 3132, and to probe its molecular gas composition, we have used the SMA to map the nebula in $^{12}$CO(2--1), as well as the 2--1 rotational transitions of CN and CO isotopologues. In this paper, we present the SMA observations of NGC 3132, and describe how these observations yield new insight into the PN's three-dimensional structure and molecular chemistry. 

\section{Observations}

We observed NGC 3132 with the Submillimeter Array (SMA) on 2023 May
16. The six operating antennas were in a compact configuration that
provided baseline lengths from 6 to 68 meters. NGC~3132 is a
challenging target for the SMA due to its southern declination
($-40\degr$) and consequent low elevations when observed from
Maunakea, requiring favorable weather conditions. For these
observations, the 225~GHz atmospheric opacity was 0.06 with very
stable phase throughout. The two dual-sideband receivers were tuned to
LO frequencies of 225.538 and 235.538~GHz. With each
receiver providing an IF range of 4-16~GHz, this setup provided
continuous spectral coverage from 209.5 to 251.5~GHz. The SWARM
digital backend provided 140~kHz channel spacing over the full
bandwidth, which corresponds to 0.18 km~s$^{-1}$ at the frequency of
the $^{12}$CO $J=2\rightarrow 1$ line (230.538~GHz). The SMA primary
beam size is $55''$ (FWHM) at this frequency. With baselines
  down to 6 meters, these SMA observations have a maximum
  recoverable scale of $\sim$27$''$. 

We observed NGC~3132 in
a small hexagonal mosaic of 7 pointings with $30''$ spacing to span
the full extent of $^{12}$CO $J=2\rightarrow 1$ emission previously
imaged with the SEST telescope (Sahai et al. 1990). The observing
sequence consisted of 2 minutes on each of the 7 mosaic pointings,
bracketed by the two calibrators J1037-295 and J1001-446. The target
was observed over the hour angle range $-$2.1 to 2.8. 

We used the MIR software package to calibrate the visibilities
following standard procedures for SMA data. The visibilities were
initially inspected manually to flag a small number of channels that
showed evidence for interference. The bandpass response was determined
from observations of the strong source 3C 279, the absolute flux scale
was set from a short observation of the asteroid Ceres (with
$\sim10$\% estimated systematic uncertainty), and time dependent
complex gains were derived and applied from observations of
J1037$-$295 (the stronger of the two gain calibrators, 1.31~Jy). 

We used the MIRIAD software package to make images, using the mosaic
option in the {\tt invert} task followed by clean deconvolution with
the {\tt mossdi} task. We imaged the $^{12}$CO $J=2\rightarrow 1$,
$^{13}$CO $J=2\rightarrow 1$, C$^{18}$O $J=2\rightarrow 1$, and CN
$N=2\rightarrow 1$ (226.875 GHz hyperfine complex) lines by generating
image cubes over velocity bins of 1.5~km~s$^{-1}$ width, chosen as a
compromise between resolving kinematic structure and signal-to-noise
ratio. Table~\ref{tbl:lines} lists the transition frequency, beam size
and position angle (PA) as obtained with robust=0 weighting, rms
channel-to-channel noise, and integrated line intensity for each of
the lines imaged. For the $^{12}$CO $J=2\rightarrow 1$ beam size
  (6.49$''$$\times$2.51$''$), the flux density to brightness
  temperature conversion is 0.71 Jy K$^{-1}$. The SMA-measured line
  $^{12}$CO $J=2\rightarrow 1$ emission morphology and line fluxes (see \S
  \ref{sec:results}) are overall consistent with those measured with
  the single-dish SEST \citep[see][their Fig.~2]{Sahai:1990fu},
  indicating that the SMA data do not suffer from significant
  interferometric flux losses. We also generated a continuum image
  using all of the bandwidth free of strong spectral lines, with an
  effective frequency of 228.7 GHz. This image has an rms noise of
  4.2 mJy~beam$^{-1}$ and shows no significant features in the central
  region of uniform noise.

\begin{table}
\begin{center}
\caption{\sc SMA Molecular Emission Line Observations of NGC 3132}
\vspace{.05in}
\label{tbl:lines}
\small
\begin{tabular}{lccccc}
\hline
Molecule (Trans.) & $\nu$ & beam size & beam PA &  rms & $I^a$\\
 & (GHz) & (arcsec$^2$) & & (mJy~beam$^{-1}$) & (Jy km s$^{-1}$) \\
\hline
\hline
$^{12}$CO ($J= 2 \rightarrow 1$) & 230.538000       & 6.5$\times$2.5 & $-12^\circ$  & 130 & 1710$\pm$6 \\
$^{13}$CO ($J=2 \rightarrow 1$)  & 220.398684  & 6.7$\times$2.6 & $-11^\circ$  & 101 & 36$\pm$9 \\
C$^{18}$O ($J=2 \rightarrow 1$) & 219.560354 &  6.9$\times$2.7 & $-8^\circ$   & 102 & $<$27 \\
CN ($N=2 \rightarrow 1$)    & 226.874781$^b$  & 6.4$\times$2.6 & $-10^\circ$  & 117 & 199$\pm$9 \\
continuum & 228.7 & 6.8$\times$2.7 & $-8^\circ$ & 4.2 & ...$^c$ \\
\hline
\end{tabular}
\end{center}

\footnotesize
{\sc Notes:} \\
a) Integrated intensity of emission and associated statistical
uncertainty; estimated systematic flux uncertainties are $\sim$10\%
(see text).\\
b) Frequency of brightest component of hyperfine complex.\\
c) 3$\sigma$ upper limit on 228.7 GHz continuum flux is $\sim$12 mJy~beam$^{-1}$.

\end{table}

\section{Results}\label{sec:results}

\begin{figure}[ht]
\begin{center}
\includegraphics[width=6.5in]{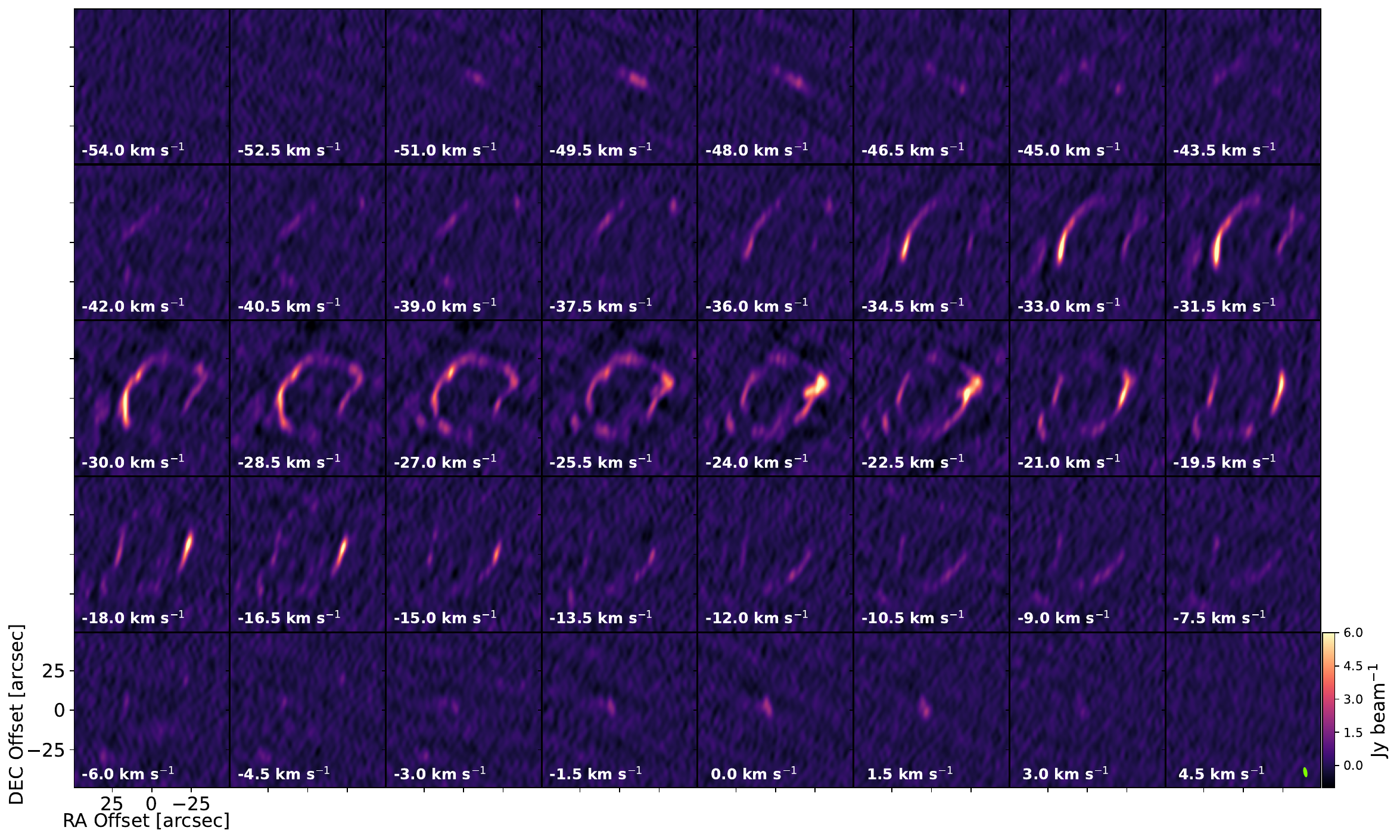}
\end{center}
\caption{SMA channel maps of $^{12}$CO(2--1) emission from NGC 3132.}
\label{fig:COchannelMaps}
\end{figure}

Channel maps obtained from the $^{12}$CO $J=2\rightarrow 1$ image cube
are presented Fig.~\ref{fig:COchannelMaps}; channel maps for $^{13}$CO
$J=2\rightarrow 1$ and CN $N=2\rightarrow 1$ are presented in Appendix
A. In Fig.~\ref{fig:mom0images}, we display velocity-integrated
(moment 0) images of $^{12}$CO $J=2\rightarrow 1$, $^{13}$CO
$J=2\rightarrow 1$, and CN $N=2\rightarrow 1$ line emission.
The corresponding respective emission line profiles,
  obtained by spatially integrating the SMA image cubes within a
  $\sim$33$''$ radius region centered on and encompassing the bright
  central molecular ring, are presented in
  Fig.~\ref{fig:spectra}.  The spectra indicate
  that the nebular systemic velocity is $\sim$$-25$ km s$^{-1}$,
  consistent with the single-dish SEST results \citep{Sahai:1990fu}. Table~\ref{tbl:lines} lists the integrated
  line intensities obtained from the spectra.
The CN radical is here detected for the first time in NGC
3132. Neither C$^{18}$O nor continuum emission were detected.

\begin{figure}[ht]
\begin{center}
\includegraphics[width=6.5in]{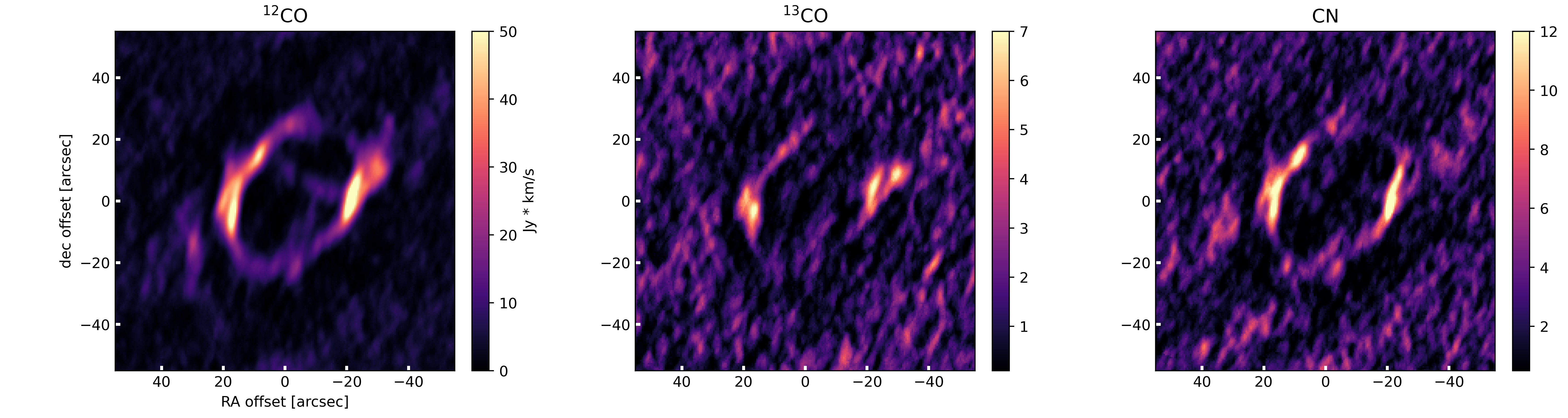}
\end{center}
\caption{From left to right, velocity-integrated (moment 0) images of $^{12}$CO(2--1), $^{13}$CO(2--1) and CN(2--1) emission, respectively, from NGC 3132 obtained from the SMA image cubes. The $^{13}$CO(2--1) and CN(2--1) moment 0 images were generated by rejecting image cube spaxels with values less than the rms noise in the data cube (see Table~\ref{tbl:lines}).}
\label{fig:mom0images}
\end{figure}

\begin{figure}[ht]
\begin{center}
\includegraphics[width=6.5in]{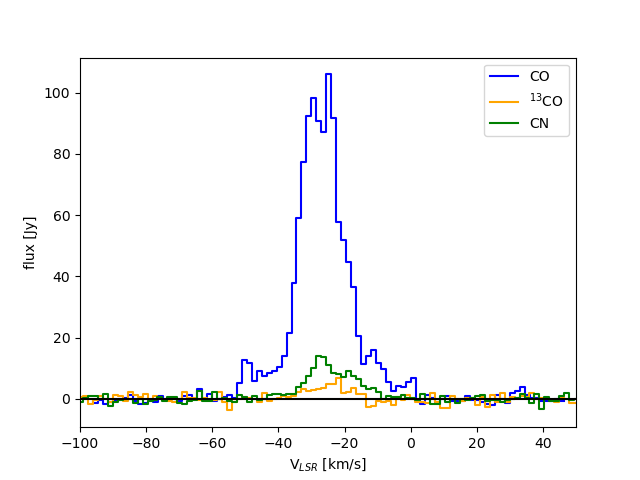}
\end{center}
\caption{Spectra of $^{12}$CO(2--1), $^{13}$CO(2--1) and CN(2--1) emission from NGC 3132 obtained by integrating the SMA image cubes within a $\sim$33$''$ radius region centered on and encompassing the bright central molecular ring. The nebular systemic velocity is $\sim$$-25$ km s$^{-1}$.}
\label{fig:spectra}
\end{figure}

It is immediately apparent from  Fig.~\ref{fig:COchannelMaps} and Fig.~\ref{fig:mom0images} that the brightest mm-wave molecular emission arises from the main ring of the nebula, as previously established by the SEST $^{12}$CO(2--1) mapping \citep{Sahai:1990fu}. However, as we describe in detail below, the $\sim$5'$''$ resolution SMA interferometric mapping elucidates various fundamental aspects of the structure of $^{12}$CO(2--1) emission that could not have been ascertained from those previous ($\sim$22$''$ resolution) single-dish SEST mapping observations.

The $^{12}$CO(2--1) emission is detected over LSR velocities ranging from $-51$ km~s$^{-1}$ to $+3$ km~s$^{-1}$ (Fig.~\ref{fig:spectra}), with the bulk of the $^{12}$CO emission arising from the bright central ring at velocities between roughly $-40$ km~s$^{-1}$ and $-5$ km~s$^{-1}$ (Fig.~\ref{fig:COchannelMaps}). The projected semimajor and semiminor axes of this main, bright, elliptical CO ring, as deduced from the velocity-integrated (moment 0) $^{12}$CO(2--1) image (Fig.~\ref{fig:mom0images}, left), are $\sim$25$''$ ($\sim$18,500 au, assuming $D = 754$ pc) and $\sim$18$''$ ($\sim$13,300 au), respectively, with the elliptical ring oriented at position angle of approximately 330$^\circ$ (as measured E from N). 

The $^{12}$CO channel maps (Fig.~\ref{fig:COchannelMaps}) furthermore demonstrate that the more highly blueshifted and redshifted features, centered at $-50$ km~s$^{-1}$ and $0$ km~s$^{-1}$, respectively (and appearing as weak ``satellite peaks'' in the $^{12}$CO(2--1) spectrum; Fig.~\ref{fig:spectra}), appear to arise from compact regions within this main ring rather than from exterior jets or ansae. The CN emission (Fig.~\ref{fig:mom0images}, right) displays the same basic (ring) emission morphology, but the signal-to-noise ratio is relatively poor and hence the detected emission is limited to velocities between roughly $-40$ km~s$^{-1}$ and $-12$ km~s$^{-1}$, while the (still weaker) $^{13}$CO emission is restricted to a still smaller velocity range (Fig.~\ref{fig:spectra}). 

\newpage

\subsection{Comparison with archival JWST H$_2$ imaging}\label{sec:SMAvsJWST}

\begin{figure}[!ht]
\begin{center}
\includegraphics[width=6.5in]{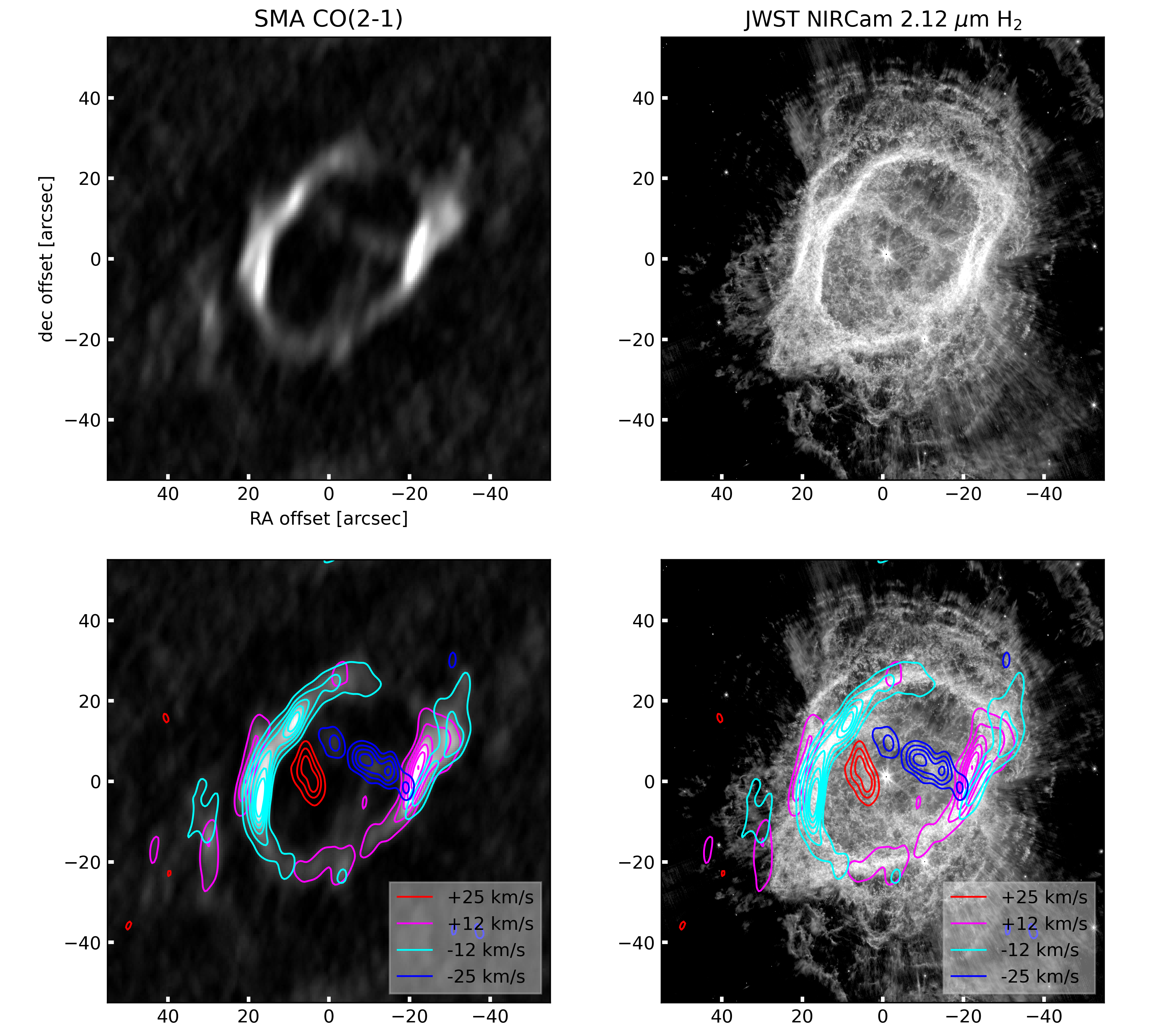}
\end{center}
\caption{Comparison of JWST/NIRCam and SMA imaging of NGC 3132. Top left: velocity-integrated (moment 0) SMA $^{12}$CO(2--1) image. Top right: JWST/NIRCam 2.12 $\mu$m H$_2$ image. Bottom left: contours of $^{12}$CO(2--1) emission integrated over selected velocity ranges overlaid on the $^{12}$CO(2--1) moment 0  image. The (four) velocity ranges, which are labeled with their central velocities with respect to systematic (i.e., $-25$, $-12$, $+12$, and $+25$ km~s$^{-1}$), correspond to LSR velocity ranges of $-52$ to $-46$ km~s$^{-1}$, $-40$ to $-25$ km~s$^{-1}$, $-25$ to $-10$ km~s$^{-1}$, and $-2$ to $+3$ km~s$^{-1}$, respectively. Bottom right: the same sets of velocity-resolved $^{12}$CO(2--1) emission contours overlaid on the JWST/NIRCam H$_2$ image. 
}
\label{fig:JWSTvsSMA2}
\end{figure}

In the upper panels of Fig.~\ref{fig:JWSTvsSMA2}, we compare the archival ERO JWST/NIRCam 2.12 $\mu$m H$_2$ image of NGC 3132 and the SMA $^{12}$CO(2--1) moment 0 image. It is apparent that there is a close morphological correspondence between the two images, despite their sharply contrasting spatial resolution ($\sim$0.2$''$ and $\sim$5$''$, respectively); the brightest near-IR H$_2$ and mm-wave $^{12}$CO are spatially coincident, and the main, bright ring appears bifurcated in the E-W direction in both images. In the lower panels of Fig.~\ref{fig:JWSTvsSMA2}, we display spectrally integrated velocity slices through the $^{12}$CO(2--1) image cube, overlaid on the SMA moment 0 image (lower left) and JWST H$_2$ image (lower right). These overlays reveal that the E-W spatial bifurcation of the bright ring has a corresponding velocity bifurcation, wherein the blueshifted (approaching) ring component is spatially offset to the west of the redshifted (receding) ring component. This resolution of the central ring into distinct spatial and velocity components suggests the ring possesses an overall cylindrical structure, and is viewed at low inclination and slightly tilted along the E--W direction with respect to the line of sight. 

Fig.~\ref{fig:JWSTvsSMA2} further demonstrates that the most highly
blueshifted and redshifted features (knots) detected in the
$^{12}$CO(2--1) mapping correspond to distinct H$_2$ filaments that
are projected within the main ring, and appear to cut across the
nebula to the northwest and southeast of the central (visual binary)
star. The northwest H$_2$ filament is evidently somewhat more
spatially extended and coherent than the southeast H$_2$ filament and,
correspondingly, the blueshifted $^{12}$CO knot is brighter and more
extended than the redshifted $^{12}$CO knot.

\subsection{Position-velocity images}\label{sec:PVimages}

\begin{figure}[ht]
\begin{center}
\includegraphics[width=2.75in]{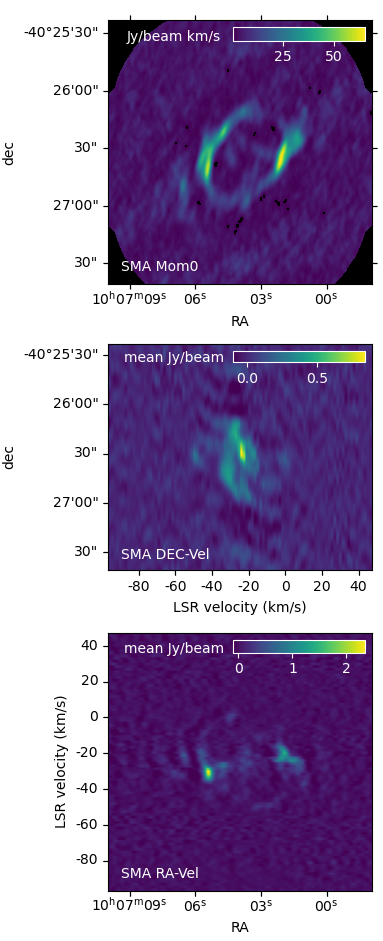}
\end{center}
\caption{Three views of the SMA $^{12}$CO(2--1) data cube. Top: the noise-clipped moment 0 image. Middle and bottom: position-velocity (P-V) images obtained by ``collapsing'' the data cube along the RA and declination axes, respectively. The position-integrated P-V images have been constructed so as to exclude P-V frames lying outside of the spectral extraction region used for Fig.~\ref{fig:spectra}.}
\label{fig:CubeMomentsSMA}
\end{figure}

\begin{figure}[ht]
\begin{center}
\includegraphics[width=6.0in]{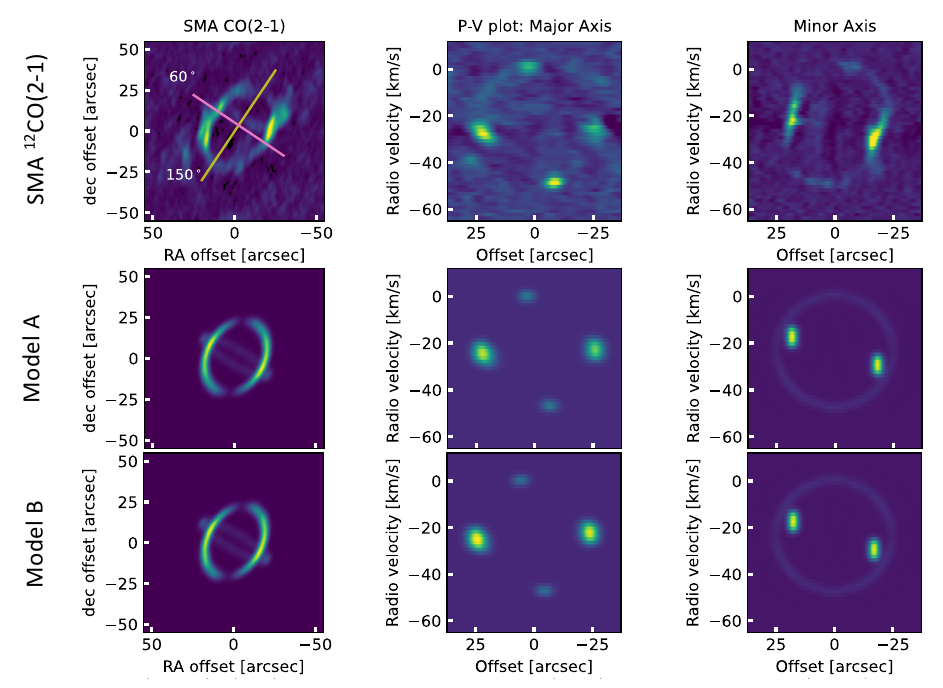}
\end{center}
\caption{{\it Top row:} Position-velocity (P-V) images of
    $^{12}$CO(2--1) emission for directions along the minor and major
    axes of the main ring of NGC 3132 (middle and right panels),
    respectively, obtained by spatially integrating slices of width
    20$''$ along position angles $60^\circ$ and $150^\circ$. These
    positions of these cuts are shown in the left panel as magenta and
    yellow lines, respectively, overlaid on the $^{12}$CO(2--1 moment
    0 image). {\it Middle row:} The corresponding moment 0 and P--V
    images for the Model A, the 2-ring model structural invoking an
    elliptical Ring 1 that is viewed at an inclination of
    20$^\circ$. {\it Bottom row:} The corresponding moment 0 and P--V
    images for Model B, the 2-ring structural model wherein Ring 1's
    dynamical age is 2.5$\times$ that of Ring 2 and is viewed at an
    inclination of 45$^\circ$. Note the similarity of the two models,
    which demonstrates the degeneracy of the model paramters. See Appendix
    B.}
\label{fig:PVcuts}
\end{figure}

In Fig.~\ref{fig:CubeMomentsSMA}, we display three views of
  the SMA $^{12}$CO(2--1) data cube, as integrated (collapsed) along each of the three cube axes. 
The velocity-integrated (moment 0) image is displayed in
the top frame, while position-velocity (P--V) images collapsed
(integrated) along the RA and declination axes are displayed in the
two panels below the moment 0 image. 
In Fig.~\ref{fig:PVcuts} (top row), we present P--V images
    obtained by spatially integrating slices of width 20$''$ through the
    SMA $^{12}$CO(2--1) data cube along position angles (PAs) of $60^\circ$ and $150^\circ$,
    corresponding to the minor and major axes of the main ring of NGC 3132, respectively.

These P--V images reveal the three-dimensional structure of the NGC
3132 molecular emission. Specifically, the RA-collapsed P--V image
(Fig.~\ref{fig:CubeMomentsSMA}, middle panel) demonstrates
that the main, bright ring seen in the near-IR H$_2$ (JWST) and
mm-wave $^{12}$CO(2--1) (SMA moment 0) imaging --- hereafter Ring 1
--- indeed has a P--V morphology consistent with an
  expanding ring. The decl.-collapsed P--V image
(Fig.~\ref{fig:CubeMomentsSMA}, bottom panel) shows that
its eastern edge is approaching, hence, tilted toward the
  observer, and its western edge is receding, hence, tilted away from
  the observer.  This P--V image and that obtained from the cut
  through the data cube along along PA $60^\circ$
  (Fig.~\ref{fig:PVcuts}, right panel) furthermore demonstrate that
  the minor axis of Ring 1 is tilted in velocity space by $\sim$10 km
  s$^{-1}$, i.e., that the line-of-sight blueshifted and redshifted (approaching and
  receding) velocities of the limbs of the ring are $\sim$5
  km s$^{-1}$. In contrast, the P--V image obtained from the
  major-axis cut through Ring 1 shows essentially no velocity tilt
  (Fig.~\ref{fig:PVcuts}, middle panel), indicating that this line
  through the ring major axis, along PA $150^\circ$, represents the
  intersection of the plane of Ring 1 with the plane of the sky.

The P--V images in Fig.~\ref{fig:CubeMomentsSMA} and
  Fig.~\ref{fig:PVcuts} (top row) also reveal the velocity coherence of the
  clumpy molecular emission structures that are seen projected within
  Ring 1. In particular, the declination-collapsed P--V image
  (Fig.~\ref{fig:CubeMomentsSMA}, bottom panel) and minor-axis P--V
  image in Fig.~\ref{fig:PVcuts} (top right panel) show that the
  high-velocity clumps seen in Fig.~\ref{fig:JWSTvsSMA2} (lower
  panels) are in fact the brightest portions of what appears to be a
  continuous ring structure in P--V space. This second, expanding ring
  of molecular gas within NGC 3132 is hereafter referred to as Ring
  2.


The SMA $^{12}$CO moment 0 images and P--V diagrams
  (Fig.~\ref{fig:PVcuts}) furthermore demonstrate that Ring 1 and Ring
  2 have very different inclinations with respect to the line of
  sight. The symmetry axis of (bright) Ring 1 is evidently viewed at
  low to intermediate inclination; specifically, its inclination is
  constrained to lie between $\sim$15$^\circ$ and $\sim$45$^\circ$.
  The lower limit on Ring 1's inclination is obtained from its $\sim$10 km s$^{-1}$
  tilt in velocity space (i.e., the $\sim$5
  km s$^{-1}$ blueshift/redshift of the ring limbs; see above) under the assumption that its expansion
  velocity is identical to that of Ring 2 ($\sim$25 km s$^{-1}$), while
  the upper limit is obtained from Ring 1's observed
  (projected) major/minor axis ratio ($\sim$1.4).

In contrast, the (fainter) Ring 2 is viewed nearly edge-on,
  and appears to be oriented such that its major axis (along position
  angle $\sim$60$^\circ$) is nearly orthogonal to that of Ring 1
  (position angle of roughly 330$^\circ$). The inclination of Ring
  2, as obtained from its major/minor axis ratio -- which we infer to
  be $\sim$4.5, based on the moment 0 and P--V images in
  Fig.~\ref{fig:PVcuts} (top row) --- is $\sim$78$^\circ$. The full
  velocity extent of Ring 2 in the P--V images is $\sim$50 km s$^{-1}$
  which, given its near edge-on orientation, implies an expansion
  velocity of $\sim$25 km s$^{-1}$. Assuming that the physical
  (linear) size of Ring 2 is similar to that of Ring 1 (i.e., radius
  of $\sim$18500 au), the dynamical age of Ring 2 is $\sim$3700 yr.

\section{Discussion}

\subsection{CO isotopologue and $^{12}$CO/CN line ratios: implications
  for progenitor mass}\label{sec:ratios} 

The $^{12}$CO(2--1)/$^{13}$CO(2--1) and $^{12}$CO(2--1)/CN(2--1)
integrated intensity line ratios measured here for NGC 3132,
$\sim$48 ($\pm$25\%) and $\sim$9 ($\pm$10\%), respectively,
(Table~\ref{tbl:lines}) are somewhat larger than measured
by \citet[][]{Bachiller1997} for the analogous (Ring-like)
  molecule-rich PNe NGC 6720 ($^{12}$CO(2--1)/$^{13}$CO(2--1) $\sim$
  22; $^{12}$CO(2--1)/CN(2--1) $\sim$ 4) and NGC 6781
  ($^{12}$CO(2--1)/$^{13}$CO(2--1) $\sim$ 17; $^{12}$CO(2--1)/CN(2--1)
  $\sim$ 6). If both the $^{12}$CO(2--1) and $^{13}$CO(2--1) lines
are optically thin, as was inferred for NGC 6720 and NGC 6781
\citep{Bachiller1997} then, neglecting chemical fractionation effects,
the spatially and spectrally integrated
$^{12}$CO(2--1)/$^{13}$CO(2--1) line ratio should directly yield a
measurement of the $^{12}$C/$^{13}$C isotope ratio within the
molecular envelope of NGC 3132. The inferred value, $^{12}$C/$^{13}$C
$\sim 50$, is consistent with the initial mass inferred for the
progenitor star \citep[i.e., $\sim$2.9 $M_\odot$;][]{DeMarco2022},
given the predictions of models of surface AGB isotope yields
\citep{KarakasLugaro2016}. Specifically, the \citet{KarakasLugaro2016}
models predict AGB surface $^{12}$C/$^{13}$C ratios in the range
$\sim$30 to $\sim$80 for progenitor masses in the range 2--4 $M_\odot$
at near-solar metallicity. The relatively weak CN emission relative to
$^{12}$CO is also consistent with the relatively low ($<$1.0) N/O
ratios expected at the surfaces of AGB stars descended from solar
metallicity progenitors in this mass range, according to the
\citet{KarakasLugaro2016} models.

\subsection{CO column densities and molecular gas mass}

To obtain $^{12}$CO column densities and (hence) estimate the total
molecular gas mass of NGC 3132 from the SMA $^{12}$CO(2--1) data, we
use the publicly available RADEX radiative transfer
code\footnote{https://home.strw.leidenuniv.nl/~moldata/radex.html}
\citep{VanderTak2007}. For an assumed molecular gas kinetic
temperature of $T_k = 30$ K \citep[see, e.g.,][]{Bachiller1997} and
H$_2$ number density of $n_{\rm H2} = 10^6$ cm$^{-3}$, RADEX
calculations indicate that the measured SMA antenna temperatures ---
which range from $\sim$0.6 Jy beam$^{-1}$ ($\sim$1.0 K) to $\sim$6.5
Jy beam$^{-1}$ ($\sim$10 K) across the individual channel maps
(Fig.~\ref{fig:COchannelMaps}) --- correspond to a range in $^{12}$CO
column densities from $N_{\rm CO} \sim$$7\times10^{14}$ cm$^{-2}$ to
$\sim$$7\times10^{15}$ cm$^{-2}$. The $^{12}$CO emission is predicted
to be optically thin ($\tau_{\rm 12CO} < 0.5$) over this domain of
$N_{\rm CO}$ for the foregoing assumptions for $T_k$ and $n_{\rm
  H2}$. For significantly smaller assumed values of $T_k$ and $n_{\rm
  H2}$, RADEX predicts that the emission would become
marginally to very optically thick, which would be inconsistent with
the relatively large measured value of $^{12}$CO(2--1)/$^{13}$CO(2--1)
$\approx$ 48 (\S~\ref{sec:ratios}).

The approximate mean of the velocity-integrated (moment 0
  image) $^{12}$CO line intensities is $\sim$30 K km s$^{-1}$. Given the
  foregoing, we hence infer that the mean integrated $^{12}$CO column
  density along a line of sight through the CO-emitting regions of the
  nebula is $\sim2\times10^{16}$ cm$^{-2}$.  For purposes of a rough
  estimate of the total molecular mass of NGC 3132, we adopt this mean
  value of $N_{\rm CO}$.  Approximating the $^{12}$CO emitting region
  of NGC 3132 as an annulus of radius 20$''$ (15,000 au) and thickness
  3$''$ (2250 au), we find that total number of CO molecules is
  $N({\rm CO}) \sim$10$^{51}$. This estimate for $N({\rm CO})$ is very
  similar to (within $\sim$30\% of) that obtained by
  \citet{Sahai:1990fu}. To convert $N({\rm CO})$ to an H$_2$ mass then
  requires an assumption for the CO abundance relative to H$_2$,
  [CO]/[H$_2$], which is a notoriously uncertain quantity
  \citep[e.g.,][and references
    therein]{Bolatto2013,Bisbas2015,Yu2017}. Adopting a plausible range
  of [CO]/[H$_2$] that is appropriate for evolved star envelopes ---
  i.e., between
  $10^{-4}$ and $10^{-5}$ \citep[see discussion in][]{Sahai:1990fu} --- we
  obtain an estimated total molecular gas mass of between $\sim$0.015
  $M_\odot$ and $\sim$0.15
  $M_\odot$ for NGC 3132.

\subsection{The structure of NGC 3132's molecular exoskeleton}

The SMA data indicate that the Southern Ring's main, bright,
molecule-rich ring (Ring 1) is indeed a ring that is viewed
  at low to intermediate inclination, as opposed to a limb-brightened
shell, in terms of its intrinsic (physical) structure
(\S~\ref{sec:results}). This conclusion, which is consistent
  with the results of the previous single-dish (SEST) CO mapping
  \citep[as interpreted by][]{Sahai:1990fu}, is supported by our
  empirical modeling of the $^{12}$CO data (Appendix B and
  Fig.~\ref{fig:PVcuts}). Evidently, the main, CO-bright reservoir of
molecular gas in NGC 3132 is largely confined to the equatorial and
lower-latitude regions of the nebula.  The SMA $^{12}$CO(2--1) mapping
results are hence consistent with those of previous surveys of H$_2$
and CO emission from PNe. As noted earlier, such molecular
emission-line surveys have established that the vast majority of
molecule-rich PNe are intrinsically bipolar in structure, with the
bulk of the molecular gas residing in equatorial tori
\citep{Kastner:1996ab,Huggins1996,Huggins2005}. The geometries of
these toroidal or ring structures are conducive to self-shielding and
dust-shielding of the molecules against the PN central star's intense,
dissociating UV irradiation \citep{Zuckerman:1988hb}. Our
confirmation that the nebula's main, bright ring is
intrinsically ring-like in structure, as opposed to a limb-brightened
shell, therefore strongly supports the hypothesis of \citet{Sahai:1990fu} that NGC 3132 is (or at least was)
  in fact a bipolar nebula with polar axis viewed at low to intermediate inclination with respect to the line of sight.

Indeed, the SMA data resolve Ring 1 into distinct spatial and velocity
components, with indications of point-symmetric structure
(Fig.~\ref{fig:JWSTvsSMA2}), suggesting that this $^{12}$CO(2--1)
emission arises from both an equatorial torus and the bases of polar
lobes. The empirical model presented in Appendix B, in which
  Ring 1 is modeled as a simple cylindrical structure, does not
  account for these features (see, e.g., the right-hand column of
  Fig.~\ref{fig:PVcuts}). Ring 1 hence may constitute the
low-latitude portions of the twin-cone (``Diabolo'') geometry that has
been proposed to explain the nebula's ionized gas morphology and
kinematics \citep{Monteiro2000,MonrealIbero2020}. The polar lobes
 have presumably expanded far enough into the ISM over
  the $\gtrsim$3000 yr (dynamical) lifetime of NGC 3132's
  ring system (see below) that any residual lobe molecular gas is now
  difficult to detect. However, it is possible that
some of the extended (halo) H$_2$ emission seen both within and
outside of the main bright molecular ring in JWST imaging arises from
such polar lobe material. If so, then this polar lobe H$_2$
  emission morphology would appear to be in conflict with the Diabolo
model, as the brightest ``halo'' H$_2$ emission in the JWST images is
more or less aligned with the major axis of Ring 1
(Fig.~\ref{fig:JWSTvsSMA2}, top right panel), whereas the Diabolo
model requires the projected lobe emission to be aligned with the
ring's minor axis \citep[see, e.g., Fig.~5 of
][]{Monteiro2000}. We note that the SMA data also reveal
  weak $^{12}$CO(2--1) emission exterior to Ring 1, in the
  form of a faint arc extending toward the ESE that has an even fainter
  potential counterpart to the WNW (see, e.g., Fig.~\ref{fig:JWSTvsSMA2}, top
  left panel). These faint molecular halo structures
  warrant confirmation and followup via deeper, wider-field CO
  mapping.

Surprisingly, the data further reveal that the nebula also appears to
harbor a second, dust-rich molecular ring (Ring 2) --- detected in
(dust) absorption, in low-excitation emission lines
\citep[][]{Monteiro2000,MonrealIbero2020,DeMarco2022}, in H$_2$
\citep{DeMarco2022}, and (now) in $^{12}$CO(2--1) --- that appears to
lie nearly perpendicular to Ring 1, at least as seen in
  projection in the SMA $^{12}$CO(2--1) moment 0 image. Under the
  assumption that Ring 2's radius is similar to the semimajor axis of
  Ring 1 (18,500 au; \S \ref{sec:results}), the measured expansion
  velocity of Ring 2 (25 km s$^{-1}$; \S \ref{sec:PVimages}) implies that the
  dynamical age of Ring 2 is $\sim$3700 yr.

Motivated by these results, we describe in Appendix B a
  simple geometrical model for the structure of NGC 3132's
  $^{12}$CO(2--1) emission regions consisting of two rings with
  sharply contrasting inclinations with respect to the line of
  sight. This simple two-ring model of NGC 3132's molecular
  exoskeleton greatly oversimplifies aspects that are readily apparent
  in the SMA observations of NGC 3132, such as the line-of-sight
  velocity extent and structure of its main, bright $^{12}$CO(2--1)
  ring (Ring 1) and the highly uneven (knotty) $^{12}$CO(2--1)
  brightness distributions of both rings. Notwithstanding its
  simplicity, this empirical model can reproduce the apparent two-ring
  structure of the $^{12}$CO(2--1) emission that is seen in
  velocity-integrated and P-V images obtained from the SMA data
  (Fig.~\ref{fig:PVcuts} and Fig.~\ref{fig:CubeMoments2}) and in
  volumetric views of the data cube itself (Fig.~\ref{fig:CubeView}),
  as well as the main features and basic shape of the SMA
  $^{12}$CO(2--1) line profile (Fig.~\ref{fig:DataVsModelSpectra}).

Based on the analysis presented in Appendix B, we
  furthermore conclude that Ring 1 either is intrinsically elliptical
  and is viewed only $\sim$20$^\circ$ from pole-on (Model A); or, if
  Ring 1 is perfectly circular and viewed more obliquely ---
  specifically, at inclination $\sim$45$^\circ$, as indicated by its
  ellipticity --- that its expansion velocity is $\sim$2.5 times
  smaller than that of Ring 2 (Model B). The parameters of these two
  alternative models (i.e., ring inclinations, expansion velocities,
  dynamical ages, and major/minor axis ratios) are listed in
  Table~\ref{tbl:params}. Synthetic moment 0 and P--V images, where the latter have
  been extracted from cuts through the Model A and Model B along PAs
  of 60$^\circ$ and 150$^\circ$, are presented in the middle and
  bottom rows of panels in Fig.~\ref{fig:PVcuts}, respectively. 

As in
  the comparisons between SMA $^{12}$CO data and models presented in
  Appendix B, Fig.~\ref{fig:PVcuts} demonstrates that Models A and B
  both well reproduce the essential aspects of the basic
  morphologies of the SMA moment 0 and P--V images (top row of
  Fig.~\ref{fig:PVcuts}), despite the fundamental differences between
  the two models. In particular, Model B requires the dynamical ages
  of the two rings to be very different --- i.e., $\sim$3700 yr (Ring
  2) vs.\ $\sim$9000 yr (Ring 1) --- whereas Model A relies on the
  assumption that their dynamical ages are identical. Furthermore, in
  Model A, the symmetry axes of the two rings are nearly orthoginal to
  one another, whereas in Model B, their inclinations differ by
  $\sim$60$^\circ$. The striking similarity of the moment 0 and P--V
  projections of these two fundamentally different two-ring model
  realizations hence emphasizes the degeneracy of the model parameters
  (ring inclinations, ellipticities, and dynamical ages).

The degeneracy
  between Models A and B can be broken via direct measurement of the
  expansion proper motion of Ring 1 from multi-epoch HST or JWST
  images, once available; such a measurement of its projected
  expansion velocity will firmly establish Ring 1's
  dynamical age. Meanwhile, for purposes of the following discussion of the potential
  shaping processes that have generated the present-day NGC 3132, we
  adopt Model A, on the basis of its relative simplicity and the
  various independent lines of evidence that favor a dynamical
  age significantly less than $\sim$9000 yr for the ionized nebula
  \citep[see, e.g.,][]{DeMarco2022}. We stress, however, that we
  cannot yet rule out Model B based on the data at hand.
  
 \newpage

\subsection{Implications for the shaping of NGC 3132 by its central star system}

The main, bright molecular ring or torus structure that dominates the
$^{12}$CO(2--1) emission from NGC 3132 (Ring 1) would appear to
closely resemble the molecular tori associated with ``classical''
pinched-waist bipolar nebulae, with perhaps the best example being NGC
6302 \citep{SantanderGarcia2017}. That nebula, like NGC 3132, harbors
a CO-bright equatorial torus with some CO emission extending into the
polar lobes. There is broad consensus that the shaping of PNe
characterized by such profound bipolar (pinched-waist plus lobe)
structures requires a close (interacting) binary companion to the
central star \citep[see, e.g.,][and references
  therein]{DeMarco:2009ab,JonesBoffin2017,Kastner2022}.  

However, the apparent presence of a second, fainter, nearly pole-on
molecular ring in NGC 3132 (Ring 2) would appear to complicate this (relatively
simple) interpretation. That is, the formation and apparent
near-simultaneous ejection of two nearly orthogonal molecular rings
implied by Model A
appears difficult to reconcile with a model of NGC 3132 as a nearly
pole-on bipolar nebula shaped by a central, interacting binary star
system. While a definitive explanation for the formation of
such a two-ring structure is beyond
the scope of this paper, we offer a general scenario here. 

If the rings indeed have very similar dynamical ages
  then --- given their similar, AGB-like expansion
velocities --- we would conclude that Ring 2 is the
remnant of the same massive ejection of molecular gas from the AGB
progenitor that generated Ring 1. It is possible that this rapid mass
loss event terminated the progenitor star's AGB evolution. The bulk of
the AGB envelope ejection was evidently focused along the equatorial
plane, forming Ring 1, but the rapid (and perhaps terminal) ejection
of the AGB star's molecule-rich envelope --- as traced in
$^{12}$CO(2--1) in the form of Ring 2 --- appears to have been overall
quasi-spherical or ellipsoidal in geometry. This would be consistent
with recent 3D morpho-kinematic modeling of long-slit spectroscopy of
[N~{\sc ii}] emission, which indicates that the central ionized gas
cavity within NGC 3132 has a prolate ellipsoidal shell structure with
its major axis oriented at $\sim$30$^\circ$ with respect to the line of sight \citep{DeMarco2022}.  

It would then remain to explain the double-ring --- as opposed to
closed ellipsoidal --- structure of the residual molecular gas that is
apparent in the SMA $^{12}$CO(2--1) data cube. One possibility is
that, after its ejection, the initially ellipsoidal molecular shell
was quickly disrupted by a rapid-fire series of misaligned jet pairs
emanating from the central multiple star system. This process might
leave only a narrow (quasi-circular or elliptical) region of the polar
lobes ``untouched,'' and this region could take the form of Ring 2,
i.e., a second molecular ring oriented nearly perpendicular to the
nebula's equatorial (molecular) torus.  

Such a scenario, though highly speculative, would be consistent with
the evidence for multiple, misaligned (possibly precessing) jet pairs
imprinted in the inferred structure of NGC 3132's central ionized
cavity and halo \cite[][]{MonrealIbero2020,DeMarco2022}. The presence
of such intermittent, wobbling (possibly precessing) jets would
strongly suggest that the mass-losing progenitor was a member of an
interacting triple (as opposed to double) star system \citep[][and
  references therein]{DeMarco2022}. As detailed in
\citet{DeMarco2022}, the likelihood that the mass-losing AGB
progenitor was a member of a hierarchical multiple system\footnote{The
A-type visual binary companion is too widely separated from the
mass-losing progenitor to have influenced its mass loss geometry, so
would represent a distant (non-interacting) fourth component of such a
multiple system.} is further supported by JWST's detection of both a
thermal IR excess from a dust disk at the central star \citep[see
  also][]{Sahai2023} and a ring or spiral pattern in the nebula's
extensive H$_2$ halo.  

The foregoing general scenario, wherein the AGB and post-AGB mass loss
leading to PN formation rapidly progresses from quasi-spherical to
highly collimated and perhaps somewhat chaotic, is of course not new;
such a model has long been invoked to explain the ongoing, rapid
structural metamorphosis of young bipolar and multipolar PNe
\citep[e.g.,][]{Sahai:1998lr,RechyGarcia2020}. We further note that
the superimposed structures observed in the young PN NGC 7027 --- a
halo ring system and equatorial molecular torus surrounding an inner
elliptical shell that has recently been punctured by a set of (three)
misaligned jet pairs \citep{MoragaBaez2023} --- would appear to make
this object a particularly close analog to NGC 3132. Indeed, NGC 3132
may offer a glimpse into the future of the disruptive processes now
underway in very young PNe such as NGC 7027. 

\section{Conclusions}

We have obtained Submillimeter Array (SMA) mapping of $^{12}$CO
$J=2\rightarrow 1$, $^{13}$CO $J=2\rightarrow 1$, and CN
$N=2\rightarrow 1$ emission from the Ring-like planetary nebula (PN)
NGC 3132. Recent JWST (Early Release Observation) infrared imaging of
NGC 3132 has revealed the structure of its H$_2$ emission region in
unprecedented detail \citep{DeMarco2022}, but provided no information
concerning its molecular gas kinematics. The velocity-resolved SMA
observations presented here, which constitute the first mm-wave
interferometric mapping of molecular line emission from the nebula,
provide additional insight into the structure of NGC 3132's molecular
envelope. Our main results and conclusions are as follows. 

\begin{itemize}

\item The bulk of the mm-wave $^{12}$CO(2--1) emission from NGC 3132
  arises from the PN's bright central ring system, with a
  velocity-integrated morphology closely resembling that of
    the brightest regions of H$_2$ emission imaged in the IR regime
  by JWST. The CN radical, a sensitive probe of N chemistry and
  photodissociation processes in PNe \citep[e.g.,][]{Bachiller1997},
  is here detected for the first time in NGC 3132. The
  velocity-integrated CN(2--1) image displays a morphology very
  similar to that of $^{12}$CO(2--1).
\item We infer $^{12}$CO(2--1)/$^{13}$CO(2--1) and
  $^{12}$CO(2--1)/CN(2--1) abundance ratios of $\sim$50 and $\sim$10,
  respectively, from the measured integrated intensity ratios. These
  abundance ratios would appear to be consistent with the initial mass
  inferred for the progenitor star \citep[i.e., $\sim$2.9
    $M_\odot$;][]{DeMarco2022}, given the predictions of models of
  surface AGB isotope yields. The mean integrated $^{12}$CO
    column density across the emitting region is found to be
    $\sim$$2\times10^{16}$ cm$^{-2}$, leading to an estimate for total
    nebular molecular (H$_2$) mass of between $\sim$0.015 $M_\odot$
    and $\sim$0.15 $M_\odot$.
\item The SMA data demonstrate that the Southern Ring's main, bright,
  molecule-rich ring (designated Ring 1) is indeed a ring
  that is viewed at low to
    intermediate inclination, as
  opposed to a limb-brightened shell, in terms of its intrinsic
  (physical) structure. It therefore appears that the main (CO-bright)
  reservoir of molecular gas in NGC 3132 is confined to the
  low-latitude regions of the nebula.  
This in turn strongly suggests that the Southern Ring is, or at least
was, in fact a bipolar nebula whose polar axis is inclined
  by $\sim$15--45$^\circ$ with respect to our line of sight.   
\item The data further reveal that the nebula also harbors a second
  molecular (CO-emitting) ring (designated Ring 2) that is
 seen projected almost orthogonally to Ring 1. We show that a simplified
  geometrical model consisting of two expanding molecular rings
  can reproduce the basic, two-ring
  structure of the $^{12}$CO(2--1) emission that is seen in
  velocity-integrated and P-V images obtained from the SMA data, as
  well as the general morphology of the spatially integrated
  $^{12}$CO(2--1) line profile. This empirical modeling exercise
  demonstrates that if Ring 1 and Ring 2 have identical expansion
  velocities ($\sim$25 km s$^{-1}$) and dynamical
  ages ($\sim$3700 yr), then Ring 1 is intrinsically elliptical and is
  viewed only $\sim$20$^\circ$ from pole-on. Alternatively, if
  Ring 1 is perfectly circular, such that its apparent ellipticity is
  entirely the result of viewing angle (inclination
  $\sim$45$^\circ$), then its expansion velocity must be $\sim$2.5 times
  smaller than -- hence, its dynamical age $\sim$2.5 times
  larger than --- that of Ring 2.
\item The apparent presence of a second, fainter, nearly edge-on
  ``twin'' to the main, bright, nearly pole-on Ring 1 would appear to
  complicate the (relatively simple) interpretation of the structure
  of NGC 3132 as a nearly pole-on bipolar nebula shaped by the
  gravitational influence of a single close companion to the
  progenitor star. 
We suggest that this apparent two-ring structure may be the remnant of
an ellipsoidal molecular envelope of AGB ejecta that has been mostly
dispersed by a series of rapid-fire but misaligned collimated outflows
or jets. Such a scenario would be consistent with the hypothesis that
the mass-losing AGB progenitor of NGC 3132 was a member of an
interacting triple star system \citep{DeMarco2022}. Detailed
simulations of the dynamical effects of such multiple-star ``toppling
jets'' systems on AGB molecular envelopes are required to test this
speculative scenario for the shaping of the molecular exoskeleton of
NGC 3132. 
\end{itemize}

Additional (sub)mm-wave (ALMA) interferometric observations of
molecular emission from NGC 3132 at higher resolution and sensitivity
are necessary. Such ALMA molecular line observations are needed both
to confirm and further elucidate the two-ring structure that is
apparent in the SMA data and, more generally, to attempt to detect and
map any cold ($\sim$30--100~K) molecular gas that lies within the
myriad faint knots and filamentary structures imaged in near-IR (hot,
$\sim$1000-3000~K) H$_2$ by JWST. Meanwhile, a second epoch of HST
and/or JWST imaging would enable measurement of the projected
expansion speed of the main, bright ring of the nebula (Ring 1), so as to
ascertain its dynamical age and thereby test the hypothesis that the
two rings mapped in CO by SMA were ejected nearly simultaneously in an
AGB-terminating mass loss episode. 

\acknowledgments{
The authors thank the anonymous refereee for many helpful comments and
suggestions. Research by J.K. and P.M.B. on the molecular content of planetary
nebulae is supported by NSF grant AST--2206033 to RIT.  R.S.'s
contribution to the research described here was carried out at the Jet
Propulsion Laboratory, California Institute of Technology, under a
contract with NASA, and funded in part by NASA via ADAP awards, and
multiple HST GO awards from the Space Telescope Science Institute. The
Submillimeter Array is a joint project between the Smithsonian
Astrophysical Observatory and the Academia Sinica Institute of
Astronomy and Astrophysics and is funded by the Smithsonian
Institution and the Academia Sinica. We recognize that Maunakea is a
culturally important site for the indigenous Hawaiian people; we are
privileged to study the cosmos from its summit. The {\it JWST} data
used in this paper can be found in MAST:
\dataset[10.17909/vhaw-9z45]{http://dx.doi.org/10.17909/vhaw-9z45}.}

\newpage

\section*{Appendix A: $^{13}$CO $J=2\rightarrow 1$ and CN
  $N=2\rightarrow 1$ Channel Maps}

\begin{figure}[ht]
\begin{center}
\includegraphics[width=5.5in]{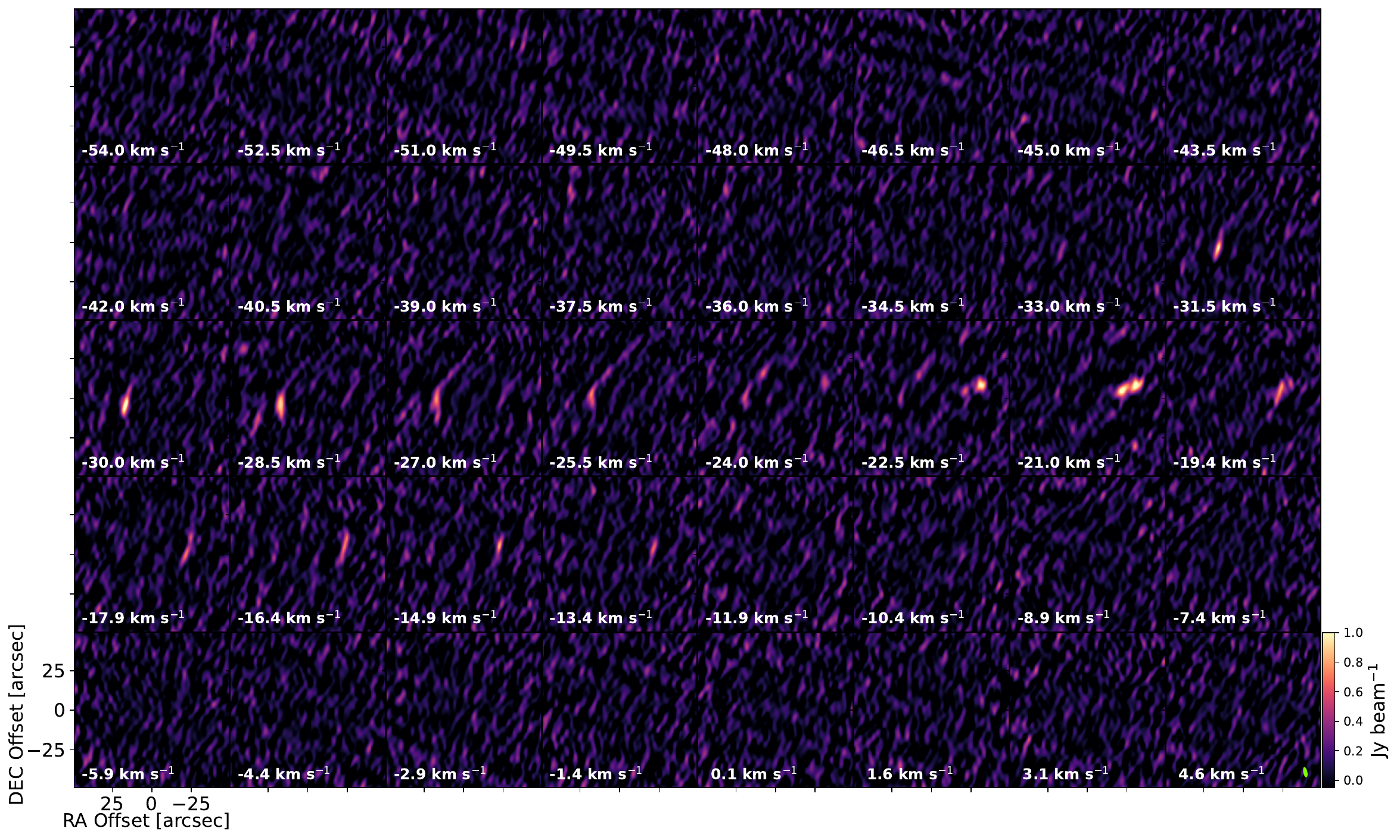}
\end{center}
\caption{SMA channel maps of $^{13}$CO(2--1) emission from NGC 3132.}
\label{fig:13COchannelMaps}
\end{figure}

\begin{figure}[ht]
\begin{center}
\includegraphics[width=5.5in]{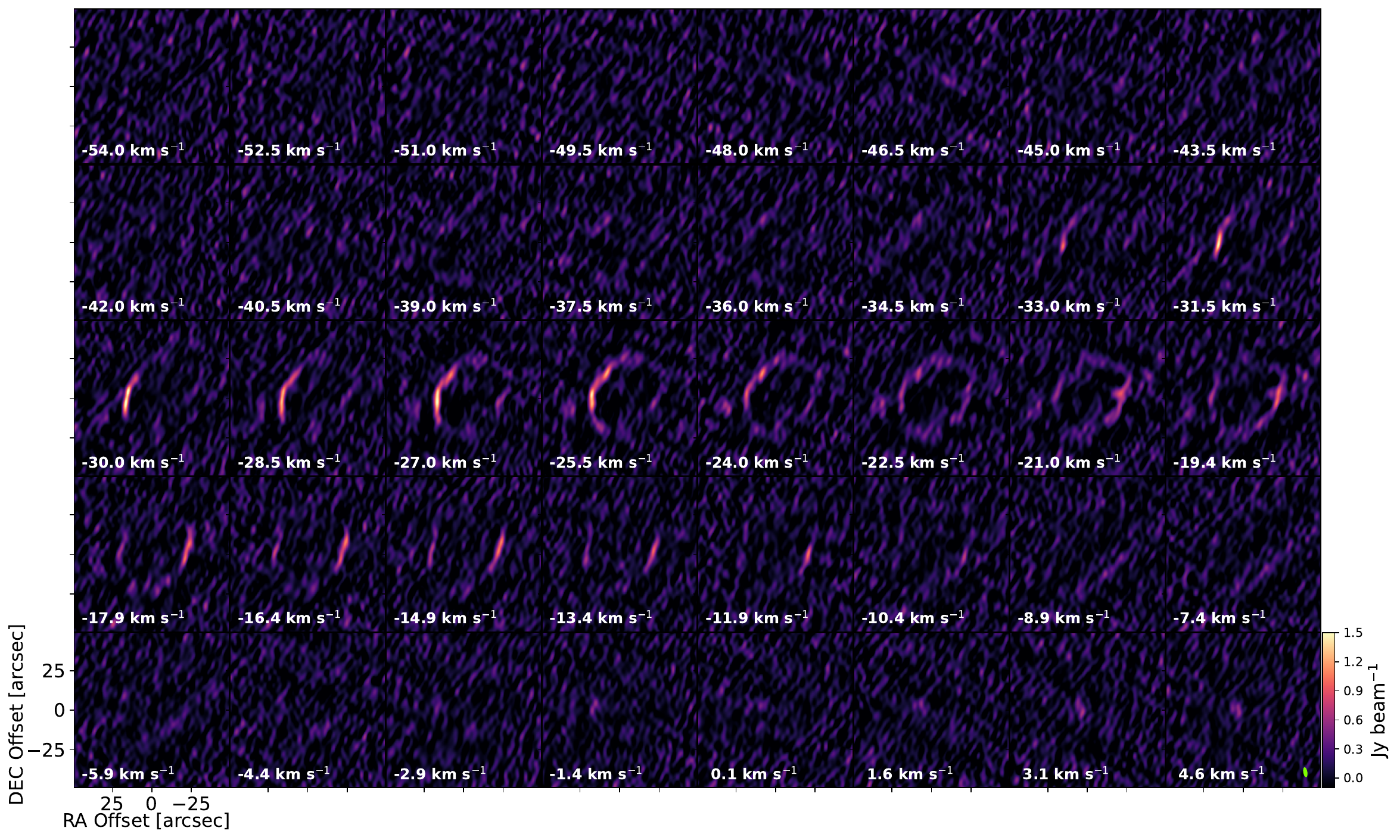}
\end{center}
\caption{SMA channel maps of CN(2--1) emission from NGC 3132.}
\label{fig:CNchannelMaps}
\end{figure}

Fig.~\ref{fig:13COchannelMaps} and Fig.~\ref{fig:CNchannelMaps} display channel maps for NGC 3132 as obtained from the SMA 
$^{13}$CO $J=2\rightarrow 1$ and CN $N=2\rightarrow 1$ (226.87478 GHz hyperfine component) line image cubes, respectively.

\newpage

\section*{Appendix B: A Simple Geometric Model}

\label{sec:model}

To interpret the $^{12}$CO emission morphology in P--V space as
revealed by the SMA data (\S~\ref{sec:PVimages}), we construct a
simple model of the main molecular gas structures within NGC 3132. We
make no attempt to model the $^{12}$CO (rotational ladder) excitation
or $^{12}$CO(2--1) photon radiative transfer; rather, the model
constructed here is purely a geometrical representation of the
$^{12}$CO(2--1) emission region within NGC 3132. Motivated and
constrained by the SMA $^{12}$CO(2--1) mapping results described in
\S~\ref{sec:PVimages}, the model we consider consists of two expanding
rings of molecular gas that are viewed at significantly different
orientations: a bright Ring 1, whose symmetry axis is viewed at
relatively low inclination with respect to the line of sight, and a
fainter Ring 2, which is viewed nearly edge-on and appears to be
oriented such that its major axis (along position angle
$\sim$60$^\circ$, measured E from N) is nearly orthogonal to that of Ring 1 (position
angle of roughly 330$^\circ$).  Both rings are
modeled in 3D cartesian coordinates $(x,y,z)$ as ellipses whose minor
and major axes lie along $x$ and $y$, respectively, and whose $z$
dimension is scaled according to the ring's expansion velocity. Ring 1
can be rotated around the $y$ axis, while Ring 2 can be rotated around the
$x$ axis. These orthogonal rotations result in ellipse dimensions along $x$ and
$z$ (Ring 1) or $y$ and $z$ (Ring 2) that are foreshortened and
modulated, respectively, by the projection effects resulting from
the rings' inclinations. 

To roughly reproduce the azimuthal asymmetry of the brighter,
lower-inclination Ring 1 as seen in the SMA moment 0 CO image
(Fig.~\ref{fig:mom0images}, left), the $(x,y,z)$ coordinates of Ring 1
are assigned intensity values ranging from 0 to 1.0, with a
$\cos{(2\phi + \phi_0)}$ dependence on azimuthal angle $\phi$ and a
rotation of $\phi_0 = 30^\circ$. The (fainter) Ring 2 is assigned
uniform intensity values of 0.125, so as to roughly match the
contribution of Ring 2 to the integrated $^{12}$CO(2--1) spectrum (see
below). Off-ring positions are assigned values of 0.

To facilitate the comparison with observations, we convolve the
resulting two-ring structure with Gaussian functions approximating the
asymmetric (6.5$''$$\times$2.5$''$) SMA beam and line spread function
(3 km s$^{-1}$, i.e., two image cube channels). To roughly approximate
the wide velocity range over which portions of Ring 1 appear to be
detected (from $\sim$$-40$ km s$^{-1}$ to $\sim$$-5$ km s$^{-1}$;
Fig.~\ref{fig:COchannelMaps}), the $z$ (velocity) dimension of Ring 1
is further convolved with a boxcar convolution kernel of width 7.5 km
s$^{-1}$. This velocity ``smearing'' is an attempt to model the
line-of-sight extent of Ring 1, and is hence distinct from the
blueshifts/redshifts due to the projected expansion
velocity of Ring 1. Finally, we reproject the model cube spaxels from
$(x,y,z)$ into the equivalent SMA data cube (RA, decl., velocity)
coordinate space and rotate the entire resulting image cube through a
position angle of 330$^\circ$, to approximate the observed (projected)
rotation of Ring 1 on the sky (\S~\ref{sec:results}).

To illustrate the output of the resulting two-ring model and the
comparison with the SMA $^{12}$CO mapping data, we first consider
two model realizations, both invoking a pair of perfectly circular rings
with identical radii (18,500 au) and expansion velocities (25 km
s$^{-1}$), hence identical dynamical ages (3700 yr). In the first
realization, the inclinations of both Rings 1 and 2 are given by
their major/minor axis ratios, as deduced from the SMA moment 0
$^{12}$CO image. These ratios are $\sim$1.4 
and $\sim$4.5 for Rings 1 and 2, respectively (see \S~\ref{sec:PVimages}), so we set
their respective model
inclinations to $\sim$45$^\circ$ and $\sim$78$^\circ$.

\begin{figure}[ht]
\begin{center}
\includegraphics[width=5.5in]{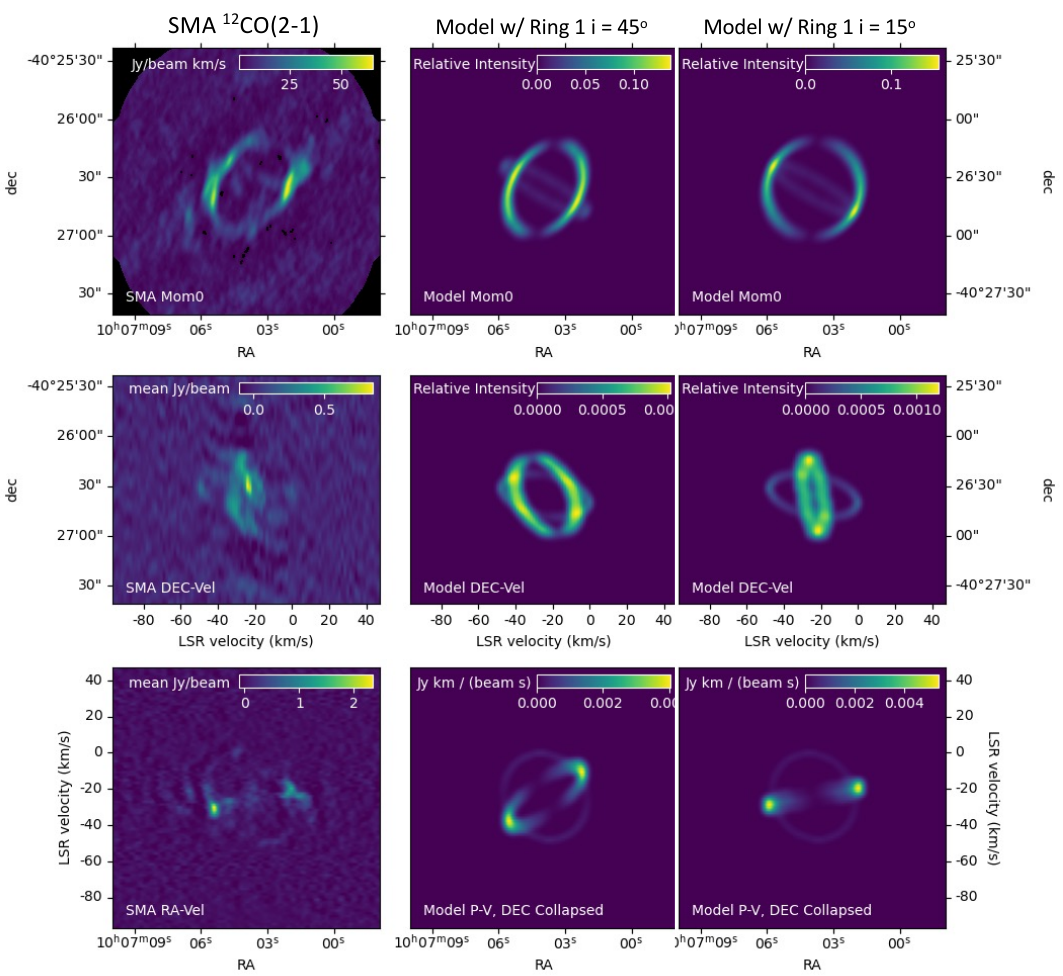}
\end{center}
\caption{{\it Left panels:} The three views of the SMA $^{12}$CO(2--1)
  data cube presented in Fig.~\ref{fig:CubeMomentsSMA}. Top to bottom:
  velocity-integrated (moment 0) image; RA-collapsed P--V image;
  decl.-collapsed P-V image. {\it Middle panels:} the corresponding
  views of the simple geometrical model of CO emission for the case of two
  circular rings with identical radii and expansion velocities and
  inclinations of 45$^\circ$ for Ring 1 and 78$^\circ$ for Ring 2. {\it
  Right panels:} the corresponding views of the same
model, but with the inclination set to 15$^\circ$ for Ring 1.}
\label{fig:CubeMoments1}
\end{figure}

The resulting structural model data cube is illustrated in the center
column panels of Fig.~\ref{fig:CubeMoments1} in the form of 
a synthetic moment 0 image and synthetic P-V images obtained by ``collapsing'' the
model image cube along the RA and declination axes. These images are
presented, alongside their observational (SMA $^{12}$CO data cube)
counterparts, in the center and left columns of
Fig.~\ref{fig:CubeMoments1}, respectively. It is readily apparent that
the synthetic moment 0 image provides a good match to the data, as
expected given that the $\sim$45$^\circ$ inclination of Ring 1 in the
model has been set by the degree of ellipticity of this Ring in the
SMA moment 0 image. However, in the RA- and decl-collapsed P--V
images, the full projected velocity extent of the model Ring 1
($\sim$40 km s$^{-1}$) is much larger than that observed ($\sim$20 km
s$^{-1}$).

In the second model realization, we attempt to qualitatively account
for this discrepancy in the velocity extent of Ring 1, by moderating
its inclination. We are guided by the observation that if the
expansion velocity of Ring 1 is identical to that measured for Ring 2,
$\sim$25 km s$^{-1}$, then the $\sim$10--12 km s$^{-1}$ tilt of Ring 1
in P--V space (Fig.~\ref{fig:PVcuts}, top right panel) would indeed imply
that Ring 1's inclination is only $\sim$15$^\circ$. We hence reduce
Ring 1's inclination accordingly, in the second model realization. The
resulting structural model data cube is illustrated in the right-hand
panels of Fig.~\ref{fig:CubeMoments1}. As expected, this model
provides an improved match to the observed RA- and decl-collapsed P--V
images, but the synthetic moment 0 image (top right-hand column panel)
fails to match the observed ellipticity of Ring 1 in the SMA moment 0
image (top right-hand column panel).

The data-model comparisons in Fig.~\ref{fig:CubeMoments1} lead us to
conclude that Ring 1 is either intrinsically elliptical in structure,
or that its expansion velocity is significantly smaller --- and hence
its dynamical age significantly larger --- than that of Ring 2.  Thus,
in an attempt to generate structural model data cubes that might more
accuractely reproduce Ring 1's morphology in all three SMA data cube
renderings --- i.e., the observed moment 0 image as well as the P--V
images --- we generate two additional, revised models corresponding to
these two possibilities. In the first revised model (Model A), Ring 1
is elliptical, with major/minor axis ratio of 1.25 and semimajor axis
of 18,500 au along the $y$ direction, and is viewed at low inclination
(20$^\circ$). The expansion velocity of Ring 1 is modulated, from a
maximum of 25 km s$^{-1}$ along the major axis to 20 km s$^{-1}$ along
the minor axis, so as to maintain a uniform dynamical age of 3700 yr
around the ring. In the second revised model (Model B), Ring 1 remains
circular, with assumed radius of 18,500 au, but its expansion velocity
is reduced to 10 km s$^{-1}$, corresponding to an increased dynamical
age of 9250 yr. The values of the fundamental parameters of Models A
and B --- ring inclination, expansion velocity, dynamical age, and
major/minor axis ratio --- are listed in Table~\ref{tbl:params}. For
simplicity, the parameters of Ring 2 are held constant for the two models.

\begin{table}
\begin{center}
\caption{\sc Two-ring CO Emission Models: Parameters}
\vspace{.05in}
\label{tbl:params}
\small
\begin{tabular}{lcccccccc}
\hline
& \multicolumn{4}{c}{Ring 1} & \multicolumn{4}{c}{Ring 2} \\
Model & $i^a$ & $V_{e}^b$ & age$^c$ & $r^d$ & $i^e$ & $V_{e}^b$ & age$^c$ & $r^d$ \\
      & ($^\circ$) & (km s$^{-1}$) & (yr) &  & ($^\circ$) & (km s$^{-1}$) & (yr) & \\
\hline
\hline
A & 20 & 25 & 3700 & 1.25 & 78 & 25 & 3700 & 1.0 \\
B & 40 & 10 & 9250 & 1.0 & 78 & 25 & 3700 & 1.0 \\
\hline
\end{tabular}
\end{center}

\footnotesize
{\sc Notes:} \\
a) Inclination of symmetry axis with respect to line of sight; b)
expansion velocity; c) dynamical age; d) major/minor axis ratio.

\end{table}

\begin{figure}[ht]
\begin{center}
\includegraphics[width=5.5in]{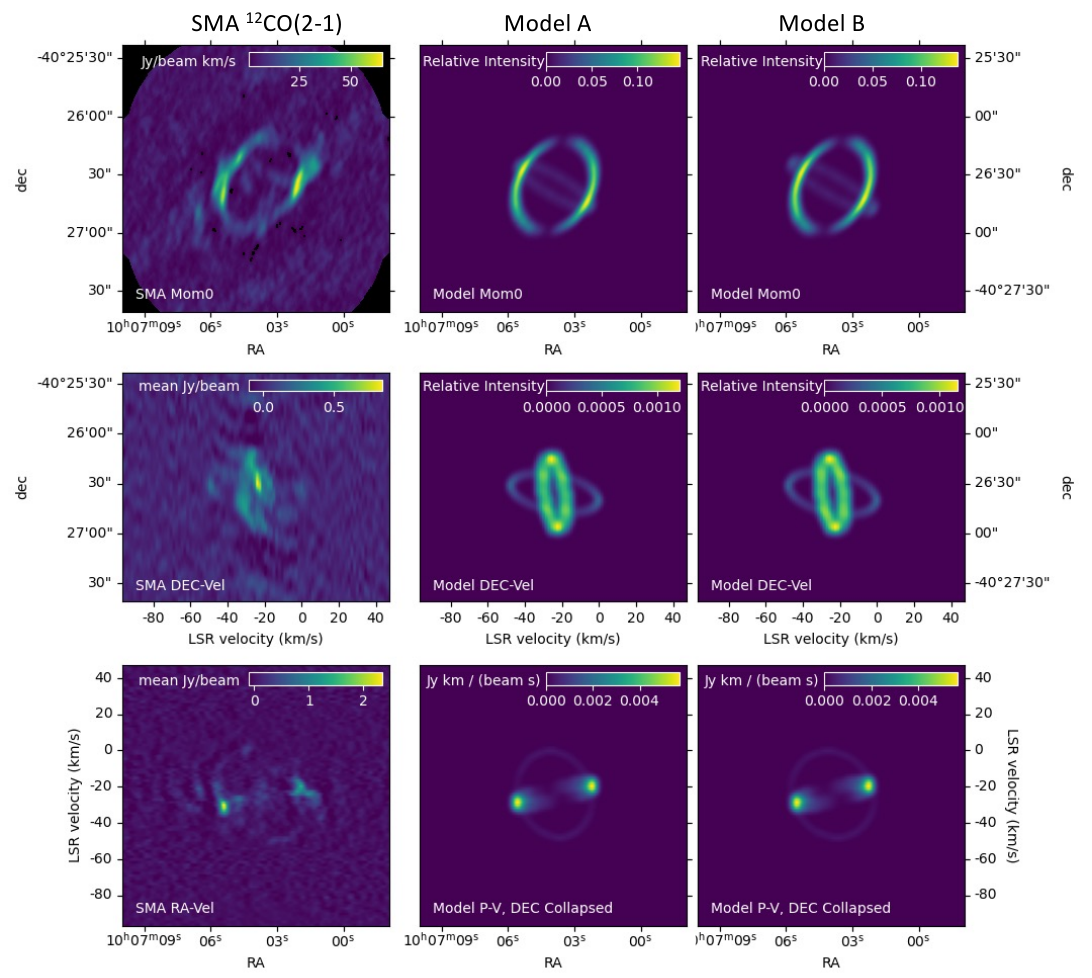}
\end{center}
\caption{As in Fig.~\ref{fig:CubeMoments1}, for two revised models
  that attempt to more accuractely reproduce Ring 1's morphology in
  all three ``collapsed'' SMA data cube renderings (again shown in the
  {\it left panels}).  {\it Middle panels:} The corresponding views of
  Model A, the simple geometrical model of CO emission for the case of an
  elliptical Ring 1 that has major/minor axis ratio 1.25 and is viewed
  at an inclination of 20$^\circ$. {\it Right panels:} The
  corresponding views of Model B, the model for which Ring 1 is
  viewed at an inclination of 45$^\circ$, as in the center column
  panels of Fig.~\ref{fig:CubeMoments1}, but its expansion velocity is
  reduced to 10 km s$^{-1}$ (2.5$\times$ smaller than that of Ring 2),
  such that its dynamical age is 9250 yr (2.5$\times$ larger than that of
  Ring 2).}
\label{fig:CubeMoments2}
\end{figure}

\begin{figure}[ht]
\begin{center}
\includegraphics[width=6.5in]{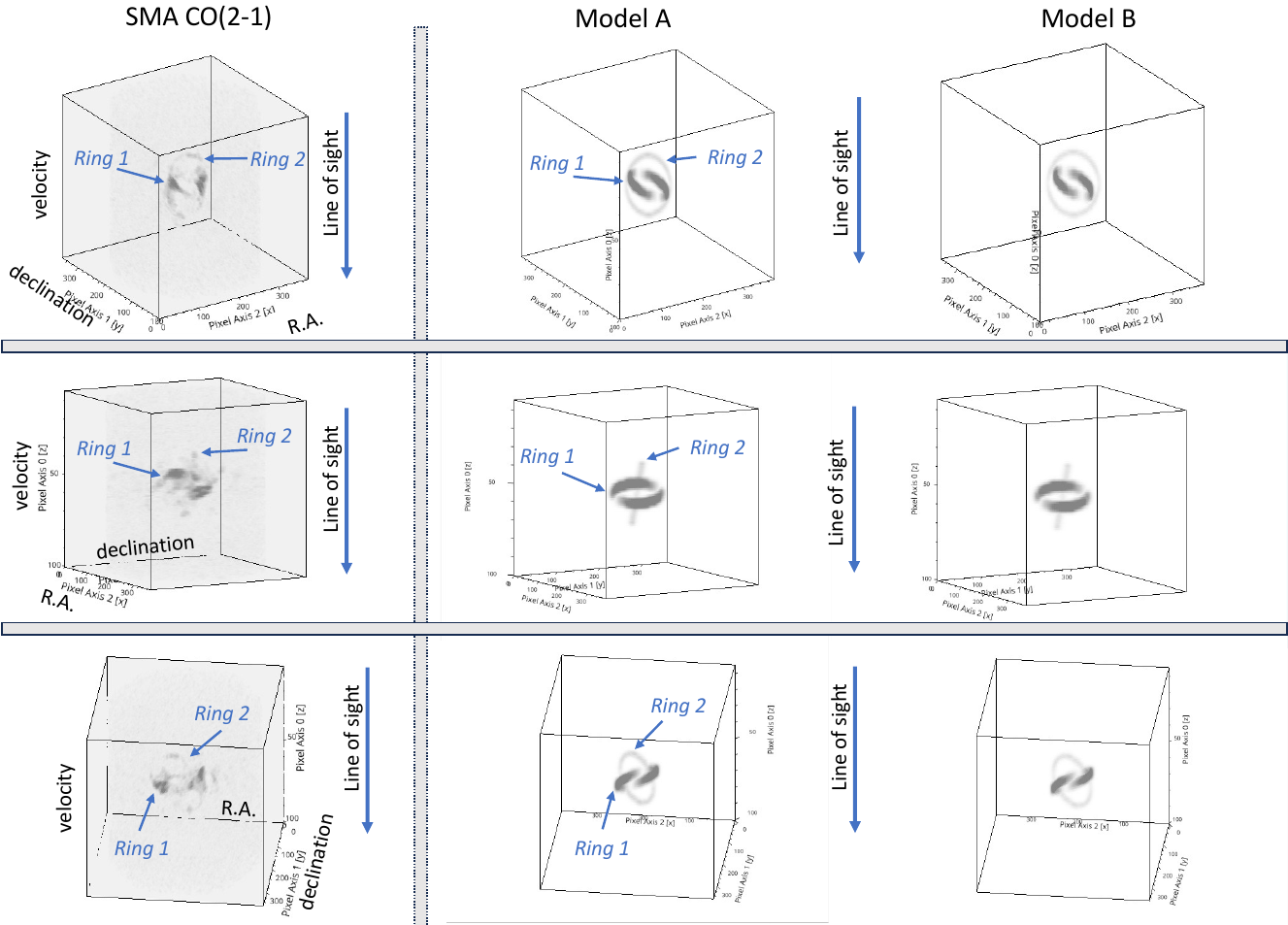}
\end{center}
\caption{{\it Left panels:} Three example volume renderings of the SMA
  $^{12}$CO(2--1) data cube, with RA, declination, and velocity as the
  $x$-axis, $y$-axis, and $z$-axis (respectively). The dimensions of
  the cube are $\sim$$180''$ along the RA, dec ($x, y$) axes and $-$96
  km s$^{-1}$ to $+$46 km s$^{-1}$ along the velocity ($z$) axis. {\it
    Middle panels:} The corresponding views of Model A. {\it
   Right panels:} The corresponding views of Model B. 
}
\label{fig:CubeView}
\end{figure}

The resulting structural model data cubes are illustrated in the
center and right-hand columns of Fig.~\ref{fig:CubeMoments2}. It is
immediately apparent that both models more closely match the data than
either of the models presented in Fig.~\ref{fig:CubeMoments1}.
Comparison of volume renderings of the SMA $^{12}$CO(2$-$1) data cube
further support the general viability of both models; one such set of
comparisons is presented in Fig.~\ref{fig:CubeView}, where we show
three example oblique views of SMA and model data cubes generated via
the {\tt Glue} software\footnote{https://glueviz.org/}. In both
Fig.~\ref{fig:CubeMoments2} and Fig.\ref{fig:CubeView}, Models A and B
are nearly indistinguishable. Evidently, the family of 2-ring models
is degenerate in terms of their possible combinations of inclination,
eccentricity, and expansion velocity (or, equivalently, assumed
dynamical age).

Despite the simplicity of the
foregoing double-ring model --- and the complexities (knots,
filaments) evident in the data that are not represented in such a
simple model --- the side-by-side comparisons in
Fig.~\ref{fig:CubeMoments2} and Fig.~\ref{fig:CubeView} demonstrate
that the two-ring model can reproduce the fundamental morphologies
apparent in the data. That is, in each of the views presented in these
Figures, the $^{12}$CO(2--1) emission appears as two intersecting
elliptical rings whose ellipticity (eccentricity) and points of
intersection are essentially functions of data cube ``viewing angle.''
Because the model does not account for opacity effects --- thereby
implicitly representing optically thin emission --- the
integrated-intensity renderings of the model in
Fig.~\ref{fig:CubeMoments2} exhibit brightness peaks at specific
locations along the two rings where each ring lies more nearly along
the axis of integration (i.e., limb brightening), or where the two
rings intersect. The $^{12}$CO(2--1) data display brightness peaks in
some of these same locations --- compare, e.g., the bottom row of panels
in Fig.~\ref{fig:CubeMoments2} --- despite the fact that the model does
not account for local density enhancements (knots) along the
rings. This supports the notion that the $^{12}$CO(2--1) emission in
Ring-like PNe like NGC 3132 is optically thin \citep{Bachiller1997}.  

\begin{figure}[ht]
\begin{center}
\includegraphics[width=3.25in]{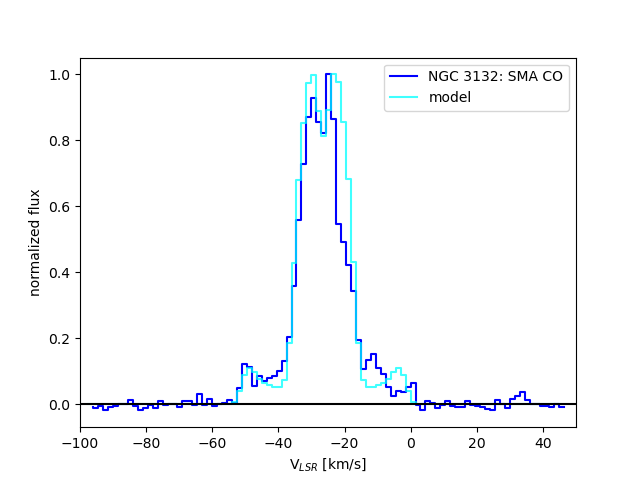}
\includegraphics[width=3.25in]{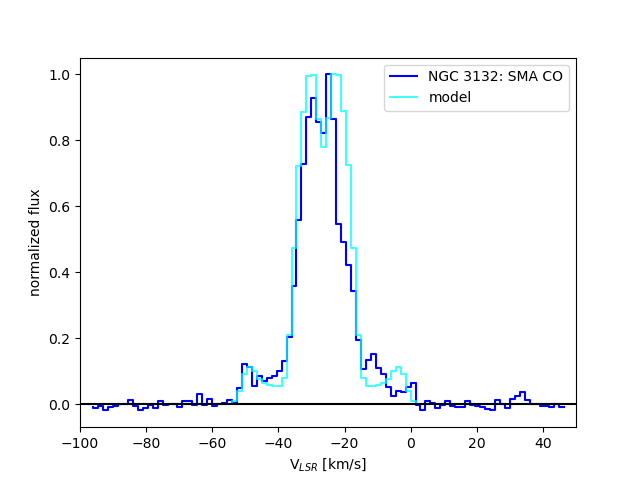}
\end{center}
\caption{Comparisons of the spatially integrated SMA $^{12}$CO(2--1)
  line profile of NGC 3132 (blue) with line profiles extracted from
  the data cubes constructed for Model A (left) and Model B (right), the simple geometrical models
  illustrated in Fig.~\ref{fig:CubeMoments2} (cyan), using the same ($\sim$33$''$ radius)
  extraction region.}
\label{fig:DataVsModelSpectra}
\end{figure}

In Fig.~\ref{fig:DataVsModelSpectra}, we present a comparison of
observed vs.\ model line profiles for Models A and B. As in the case
of the moment 0 and P--V images and data cube renderings presented in
Fig.~\ref{fig:CubeMoments2} and Fig.~\ref{fig:CubeView}, the model
line profiles for the two models (shown in cyan in
Fig.~\ref{fig:DataVsModelSpectra}) are nearly indistinguishable. This
Figure demonstrates that, in both cases, the simple 2-ring model
described here can well reproduce the observed width and double-peaked
profile of the bright line core, which is dominated by Ring 1.  Both
models also well reproduce the blueshifted ``satellite peak'' at
$\sim$50 km~s$^{-1}$ generated by Ring 2. However, the model fails to
reproduce the detailed shapes of the line wings and the redshifted
satellite peak at $\sim$0 km s$^{-1}$; the latter mismatch can be
attributed to the patchy/knotty nature of the rearward (redshifted)
side of Ring 2 (see Fig.~\ref{fig:JWSTvsSMA2}, bottom frames).

\newpage




\end{document}